\documentclass[10pt]{iopart}
\usepackage{iopams}  
\usepackage[normalem]{ulem}
\usepackage{wasysym}
\usepackage{color}
\usepackage{graphicx}
\usepackage{dcolumn}    % Align table columns on decimal point
\usepackage{bigstrut}
\usepackage{bm}
\usepackage{CJK}
\usepackage[pdfstartview=FitH,
            CJKbookmarks=true,
            bookmarksnumbered=true,
            bookmarksopen=true,
            colorlinks,
            pdfborder=001,
            linkcolor=blue,
            anchorcolor=blue,
            citecolor=blue
            ]{hyperref}
\begin{document}
\begin{CJK}{UTF8}{gbsn}
\title{Extended Lipkin-Meshkov-Glick Hamiltonian}

\author{R. Romano$^{1}$, X. Roca-Maza$^{1,2}$, G. Col\`o$^{1,2}$, Shihang Shen (申时行)$^{1,2}$}
\address{$^{1}$Dipartimento di Fisica ``Aldo Pontremoli'', Universit\`a degli Studi di Milano, 20133 Milano, Italy}
\address{$^{2}$INFN,  Sezione di Milano, 20133 Milano, Italy}

\ead{xavier.roca.maza@mi.infn.it}

%=============================================================
% ABSTRACT
%=============================================================
\begin{abstract}
  The Lipkin-Meshkov-Glick (LMG) model was devised to test the validity of different approximate formalisms to treat many-particle systems. The model was constructed to be exactly solvable and yet non-trivial, in order to capture some of the main features of real physical systems. In the present contribution, we explicitly review the fact that different many-body approximations commonly used in different fields in physics clearly fail to describe the exact LMG solution. With similar assumptions as those adopted for the LMG model, we propose a new Hamiltonian based on a general two-body interaction. The new model (Extended LMG) is not only more general than the original LMG model and, therefore, with a potentially larger spectrum of applicability, but also the physics behind its exact solution can be much better captured by common many-body approximations. At the basis of this improvement lies a new term in the Hamiltonian that depends on the number of constituents and polarizes the system; the associated symmetry breaking is discussed, together with some implications for the study of more realistic systems.  
\end{abstract}

\vspace{2pc}
\noindent{\it Keywords}: Solvable models, Symmetry breaking, Nuclear structure.

\submitto{\jpg}

%\ioptwocol

\newpage

A complete microscopic study of quantum many-body systems with realistic Hamiltonians is the central problem of different fields in physics such as atomic physics, condensed matter, or nuclear physics among others \cite{wiringa1993, sapirstein1998, onida2002, bloch2008, amico2008,katsnelson2008}. The systems under study are in general very complicated to tackle and different many-body approximations have been proposed along the years in order to understand the diverse phenomenology: from pairing in superconductors or superfluidity in ${}^{3}$He to collective states in nuclei. However, to test, compare and better understand state-of-the-art formalisms may become unpractical, especially when originating from different fields. One way forward is to simplify the problem under study by proposing a more simple albeit non-trivial Hamiltonian. That is, an Hamiltonian that contains some of the relevant features of the physical system under study. Then, testing and learning from the very complex many-body techniques available in the literature may become a manageable alternative \cite{Lipkin, Ring_Schuck,simple_model1,simple_model2}. 

With this aim, the LMG model was proposed \cite{Lipkin}. Such model considers a system of $N$ fermions distributed on two levels, each of them $M$-fold degenerate and separated by an energy $\varepsilon$. Each state is described by two quantum numbers: $\sigma$ specifies the level 
(we will refer to the upper and lower levels by using $+$ and $-$, respectively) and $p$ specifies the particular degenerate state within a given level. In this schematic model, fermions interact by a monopole interaction that does not change the $p$ quantum number. The interaction has two channels. The first scatters pairs of particles in the same level to the other level while the other scatters one particle to the upper level and, at the same time, another to the lower level. Since each particle has only two possible states, the model can be also understood as a system of spins. In absence of interaction, the model will predict all spins aligned along the same direction while other more complex configurations will be only favoured when the interaction is switched on. As it can be easily understood from this analogy, the power of the LMG model arises from the fact that it mimics some features of different physical many-body systems and, in addition, it admits an exact solution: using a quasi-spin formulation, the Hamiltonian can be written in terms of the operators that generate the SU(2) algebra.

To date the LMG model \cite{Lipkin} has been applied and extended in a variety of fields and for the study of different phenomena. An introduction to the LMG model has been given in the context of the many-body nuclear problem in Ref. \cite{Ring_Schuck}. The LMG has been studied at finite temperature within the mean-field approximation \cite{blin1996}, applied to the study of quantum phase transitions \cite{quan2007,Solinas_ribeiro}, spontaneous symmetry breaking \cite{jahn1937,bersuker2006,reinhard1984,ham1965,millis1996,kugel2018,vendrell2018,liu2019}, finite size effects \cite{dusuel2004,dusuel2005,ribeiro2008,huang2018}, long-range interacting spin-chains \cite{Silva_LMG}, quantum metrology \cite{Paris}, optical cavity QED \cite{morrison2008}, Bose-Einstein condensation \cite{zibold2010}, quantum spin squeezing \cite{ma2011,zhang2017} or quantum entanglement \cite{orus2008,zhang2013}, among others \cite{ma2017,ma2017_2}. It has been extended to three level systems \cite{Hagino_bertsch,grasso2000} and used to test different many-body approaches such as the Random Phase Approximation (RPA) \cite{Co_leo_lipkin, Marinelli}, the coupled-cluster method \cite{CC_lipkin} or density functional theory \cite{Lacroix_dft_lipkin}. Nevertheless, the modification of the Hamiltonian that we discuss below, and that is called for by the comparison with a full two-body Hamiltonian, has not been previously studied.  

For the two level system described above, the most general Hamiltonian that does not change the quantum number $p$, written in second quantization, reads
\begin{eqnarray}
  \mathcal{H}&=&\frac{1}{2}\varepsilon\sum_{p\sigma}\sigma a^\dag_{p\sigma}a_{p\sigma}\nonumber\\ &+& \frac{1}{2}\sum_{\sigma_1,\sigma_2,\sigma_3,\sigma_4}\sum_{p,p^\prime} \mathcal{V}_{\sigma_1\sigma_2,\sigma_3\sigma_4} a^\dag_{p\sigma_1}a^\dag_{p^\prime\sigma_2}a_{p^\prime\sigma_4}a_{p\sigma_3}.
\label{eq1}  
\end{eqnarray}
Here, for convenience, we have introduced a shorthand notation for the matrix elements $\mathcal{V}_{\sigma_1\sigma_2,\sigma_3\sigma_4} \equiv \langle p\sigma_1, p^\prime\sigma_2\vert \mathcal{V}\vert p\sigma_3, p^\prime\sigma_4\rangle$. Expanding the sums in Eq.~(\ref{eq1}), and with very mild assumptions (see below), one can arrive at the following expression for the Hamiltonian:  
\begin{eqnarray}
  \mathcal{H}&=&\varepsilon J_z - \frac{V}{2}(J_+^2+J_-^2) - \frac{W}{2}(J_+J_-+J_-J_+)\nonumber \\ &&- G(J_++J_-)(N-1) + \frac{W}{2}N-\frac{F}{2}N(N-1) \ . 
\label{eq2}  
\end{eqnarray}
The various steps to derive Eq. (\ref{eq2}) are described in the Appendix.
We have defined the quasi-spin operator $\bm J$ and the particle number operator $N$ as follows,
\begin{eqnarray}
  J_z=\frac{1}{2}\sum_{p\sigma} \sigma a^\dag_{p\sigma}a_{p\sigma}, &~~~~&N=\sum_{p\sigma} a^\dag_{p\sigma}a_{p\sigma}, \nonumber \\  
  J_+=\sum_{p} a^\dag_{p+}a_{p-},&~~~&J_-=\sum_{p} a^\dag_{p-}a_{p+} \ .
\label{eq3}  
\end{eqnarray}
The components of the quasi-spin operator $\bm J$ follow the usual SU(2) algebra, namely $[J_+,J_-]=2J_z$ and $[J_z,J_\pm]=\pm J_\pm$. The particle number operator commutes with all of them. Hence, the Hamiltonian (\ref{eq2}) can be solved exactly using the angular momentum representation. The coupling constants $V, W, G$ and $F$ in Eq. (\ref{eq2}) are defined in terms of the matrix elements in Eq.~(\ref{eq1}) as follows,
\begin{eqnarray}
& -V\equiv \mathcal{V}_{++,--}  &~~~~ -F\equiv \mathcal{V}_{+-,+-}=\mathcal{V}_{++,++}=\mathcal{V}_{--,--} \nonumber\\
& -W\equiv \mathcal{V}_{+-,-+}  &~~~~ -G\equiv \mathcal{V}_{++,-+} =\mathcal{V}_{--,+-} \ . 
\label{eq4}  
\end{eqnarray}
The equalities (\ref{eq4}) between the matrix 
elements defining $F$ and $G$ in our model are representative of physical cases in which $\mathcal{V}_{++,++}$ 
and $\mathcal{V}_{--,--}$, as well as $\mathcal{V}_{++,-+}$ and $\mathcal{V}_{--,+-}$, can be expected to be of the same order. 
Hence, one may also assume $\mathcal{V}_{+-,+-}=(\mathcal{V}_{++,++}+\mathcal{V}_{--,--})/2$. These considerations allow us 
a factoring of the terms $\sum_{p} a^\dag_{p+}a_{p+}$ and $\sum_{p} a^\dag_{p-}a_{p-}$ that otherwise will remain 
separated and form the number operator, and this enables a compact and convenient representation of the Hamiltonian (\ref{eq2}) 
(see the Appendix for the details). 

For systems with a fixed number of particles, the last two terms in Eq.~(\ref{eq2}) produce just a constant shift in the energy and, thus, can be dropped without losing generality. Then, the term in $G$ in the Hamiltonian (\ref{eq2}) becomes a one-body term. Without it, 
one immediately recovers the original LMG model proposed in Ref. \cite{Lipkin}. In the present work we propose instead to keep the more general formulation of the Hamiltonian within the two-level assumption of the LMG model, that is, 
\begin{eqnarray}
  \mathcal{H}&=&\varepsilon J_z - \frac{V}{2}(J_+^2+J_-^2) - \frac{W}{2}(J_+J_-+J_-J_+)\nonumber \\ &&- G(J_++J_-)(N-1) \ .
\label{eq5}  
\end{eqnarray}
We will call this Extended LMG (ELMG) Hamiltonian. The new term proportional to $G(N-1)$ scatters one particle upward or downward. It is important to note that this term:
\begin{itemize}
\item[i)] has its origin in the general form of the two-body interaction [cf. Eq. (\ref{eq1})], and yet it has been turned into a one-body term;
\item[ii)] produces a fundamental difference in the energy dependence on the particle number, when compared to the LMG model; and,
\item[iii)] introduces an explicit symmetry breaking.
\end{itemize}
In order to clarify the last point, we recall 
that in the LMG Hamiltonian the unperturbed term depends on $J_z$, 
the term in $W$ 
is proportional to $J^2-J_z^2$, and the term in $V$ is proportional 
to $J_x^2-J_y^2$. 
These facts are reflected in the symmetries displayed by
the original LMG model, that can be found in Sec. 2 of Ref. \cite{Lipkin}.
If $V=0$, $J^2$ and $J_z$ are conserved. If $W=0$  
a discrete symmetry exists, namely a rotation of $\pi$ about 
an axis in the xy-plane at an angle of $\frac{\pi}{4}$ to the x and y axes 
changes $H$ into $-H$. Such limiting cases do not exist in the ELMG model
any longer. The only symmetry which is common to the LMG and ELMG  Hamiltonians 
is that they commute with $J^2$ for all values of the parameters.

To the best of our knowledge, a study of the symmetry breaking term that we have just introduced, motivated by the comparison with a general two-body Hamiltonian, has not been yet performed. Within the context of the three-level Lipkin model, the authors of Ref. \cite{Bertolli} have introduced a different kind of symmetry-breaking term, that actually breaks the degeneracy of the levels but does not change the structure of the interaction terms. In the case at hand, we should at this point notice that it would be possible to diagonalise the one-body part of the ELMG Hamiltonian (\ref{eq5}). 

Let us introduce a new set of fermion operators $b^\dag_{p,\sigma}$, that are related to 
$a^\dag_{p,\sigma}$ by a unitary transformation, namely
\begin{equation}
\left(\begin{array}{c}
  b^\dag_{p,+}\\
  b^\dag_{p,-}
\end{array}\right)
=
\left(\begin{array}{cc}
 \cos\frac{\alpha}{2} & -\sin\frac{\alpha}{2}\\
 \sin\frac{\alpha}{2} &  \cos\frac{\alpha}{2}
\end{array}\right)
\left(\begin{array}{c}
  a^\dag_{p,+}\\
  a^\dag_{p,-}
\end{array}\right). 
\label{eq6a}
\end{equation}
Without loss of generality, $\alpha$ can be restricted to the interval 
$-\pi < \alpha \le \pi$. This transformation, in general, turns the SU(2) set of operators 
into a new one. 
It can be easily shown to 
generate a diagonal one-body Hamiltonian out of (\ref{eq5}), if 
one picks up the specific value $\alpha = {\rm artg}\frac{2G(N-1)}{\varepsilon}$. 
Under such particular transformation, the two-body sector of the Hamiltonian changes in a 
non-trivial manner. Then, we can claim that the model associated with the new ELMG Hamiltonian (\ref{eq5}) cannot be trivially mapped onto either the original LMG model or the model introduced in Ref. \cite{Bertolli}.

\begin{figure}[t!]
\vspace{5mm}  
\includegraphics[width=\linewidth,clip=true]{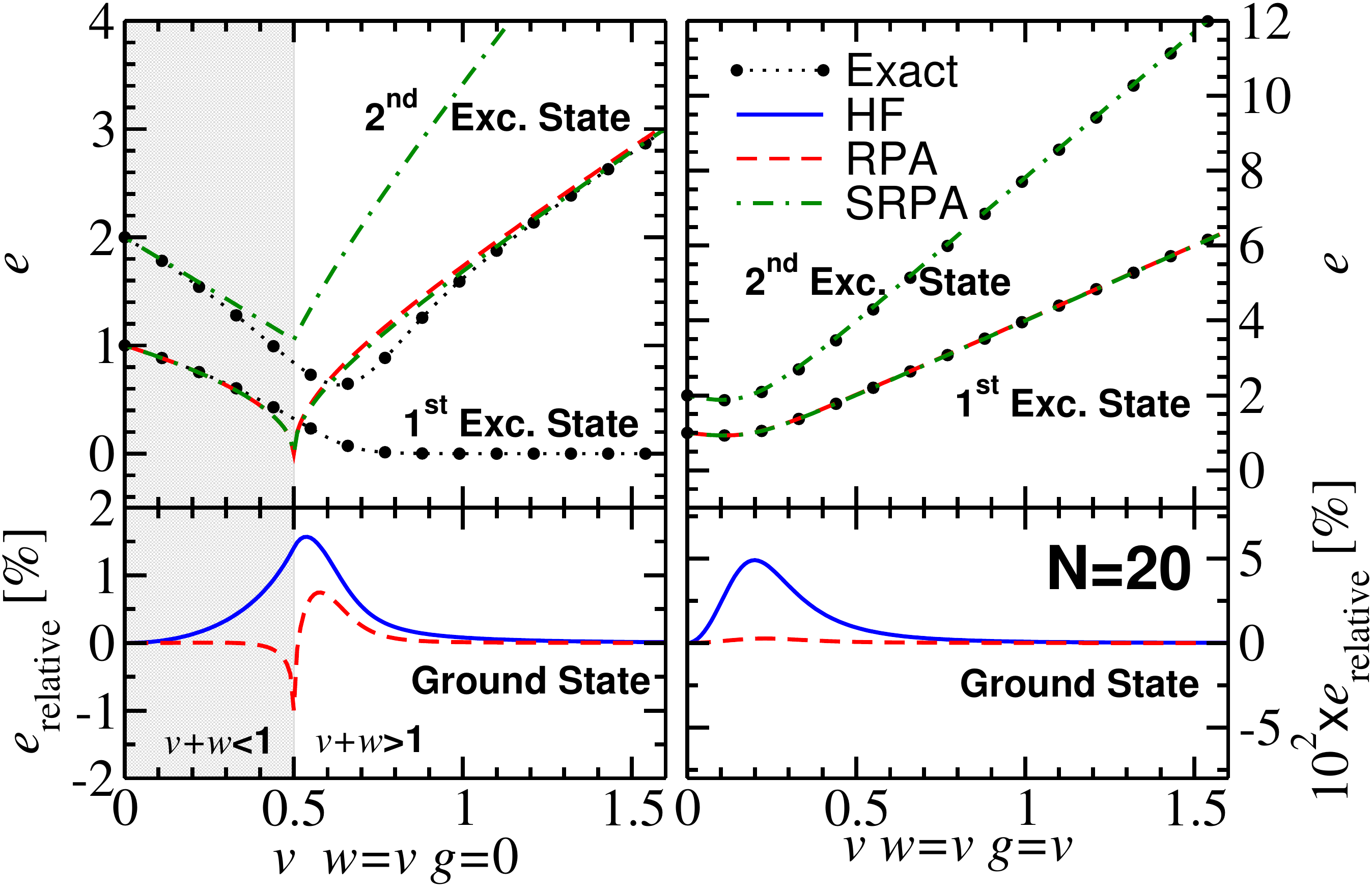}
\caption{Relative energy of the ground state with respect to the exact result [$\%$], and energy $e\equiv E/\varepsilon$ of the first and second excited states, for a system with $N=20$ particles as a function of the model parameters $v\equiv V(N-1)/\varepsilon$, $w\equiv W(N-1)/\varepsilon$ and $g\equiv G(N-1)/\varepsilon$. The exact solution is compared to HF, RPA and SRPA (this latter only for excited states). The left panel corresponds to the LMG  model for $v=w$ and the right panel to the new ELMG  model for $v=w=g$.}
\label{fig1}
\end{figure}

We now move to the analysis of the results that can be obtained by solving the new ELMG model, by comparing with well-known LMG results. In Fig.~\ref{fig1}, we show some exact and approximate results for the LMG model (left panels) and for the ELMG model (right panels). For details on the exact solution and some approximate solutions of the LMG model, we refer the reader to Refs. \cite{Lipkin,Co_leo_lipkin}. The methodology to solve the ELMG model is fully analogous to that for the LMG model. In the figure, we explore the behaviour of the ground state energy, as well as of the lowest excited state energies, as a function of the model parameters, by comparing exact results for a system with $N=20$ particles with three different approximations of common use in many-body physics \cite{Ring_Schuck}: the Hartree-Fock (HF) approximation, the 1-particle 1-hole Random Phase Approximation (RPA), and the 2-particle 2-hole RPA or second RPA (SRPA). In the figure, we have redefined the coupling constants as $v\equiv V(N-1)/\varepsilon$, $w\equiv W(N-1)/\varepsilon$ and $g\equiv G(N-1)/\varepsilon$, and the total energy as $e\equiv E/\varepsilon$ where $E$ is the total energy associated with the solution of the Hamiltonian. We have chosen to show the results for some specific values of the parameters, but the general features displayed in Fig. \ref{fig1} lead to similar conclusions for other choices. 

\begin{figure}[t!]
\vspace{5mm}  
\includegraphics[width=\linewidth,clip=true]{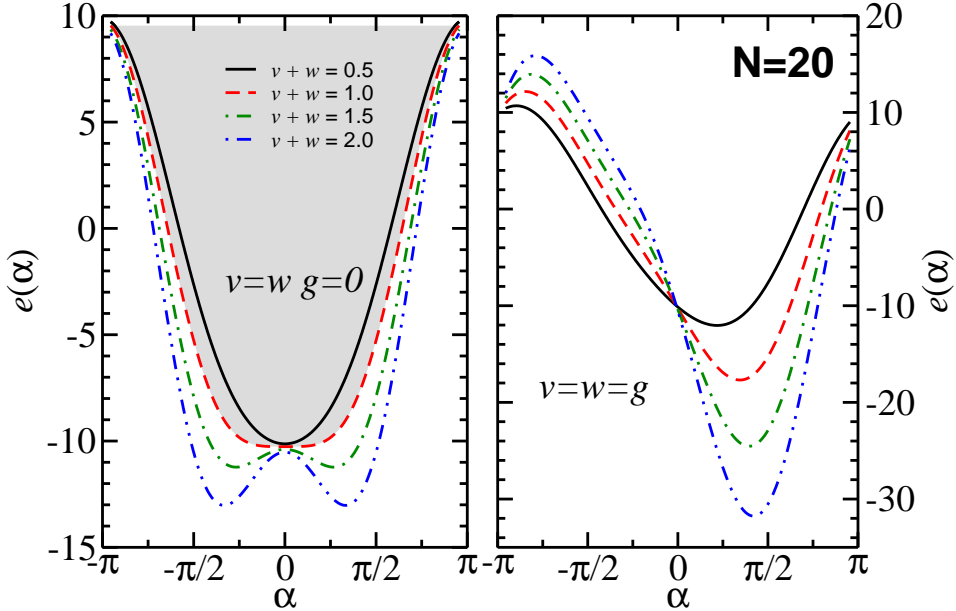}
\caption{Energy $e(\alpha)\equiv E(\alpha) / \varepsilon$ (defined in 
the text) as a function of the variational parameter $\alpha$, for different values of the coupling constants, as predicted by the LMG model (left panel) and the ELMG model (right panel).}
\label{fig2}
\end{figure}

In the left panels, the results for the LMG model are shown. Two regions are highlighted since the LMG model predicts the existence of two different HF ground states depending on the value of the parameters $v$ and $w$: $v+w<1$ produces a non-degenerate, and $v+w>1$ a degenerate, ground state (cf. Fig. \ref{fig2}). In both regions, the HF ground state (left lower panel) deviates from the exact solution by a few \% with a maximum deviation around the border between the two regions ($v+w\approx 1$). A similar trend is also found for the RPA ground state, although it is more accurate than the HF prediction. Regarding excited states, the approximate solutions of the LMG model gives a reasonable description of the exact one in the first region $v+w<1$, while they completely fail in the second region. 

In the right panels, we show the results for the ELMG model. In 
this case, the HF ground state is unique (cf. Fig. \ref{fig2}) and no regions are highlighted.
From the upper panel one can see that the excitation spectrum is different than in the case
of LMG but, also, that the accuracy of approximate many-body theories is remarkable, at 
variance with the case of LMG. 
For the ground state, the deviation of HF and RPA is below 0.05\% (note the different 
vertical scale in the lower-right and lower-left panels), while the approximate excited states 
follow closely the exact solution. 
It is interesting, though, to note that for $v=0.2$ the exact ground-state 
displays the largest difference with the HF ground-state. In fact, we 
have checked that RPA ground-state
correlations reach a maximum for that value of $v$. 

Therefore, it is clear from Fig. \ref{fig1} that the addition of the term in the Hamiltonian proportional to $G$ produces a fundamental difference in the ground state of the system when solved within the simplest many-body approximation, that is, within the HF approach. The HF ground state can be used as a basis for higher-order approximations, like RPA and SRPA, in a conceptually similar way in which it serves as a basis for second order --and higher orders-- many-body perturbation theory. On this regard, we have noticed by performing SRPA calculations that the LMG model lacks contributions form the coupling of 1$p$-1$h$ to 2$p$-2$h$ states, which is a consequence of the simplicity of the Hamiltonian. This has been already emphasised in point i) below Eq. (\ref{eq5}). As a consequence, many-body methods beyond the RPA should not be expected to be accurate in the LMG model. This problem is overcome by the new ELMG Hamiltonian. In short, we can stress that this new Hamiltonian is more general than the original LMG Hamiltonian and, therefore, has a potentially larger spectrum of applicability; moreover, common many-body approximations capture much better the physics behind the exact solution of the ELMG model.

Let us now inspect in some detail the HF ground state in the two models. 
A Slater determinant has the general form 
$\vert \alpha \rangle \equiv \Pi_{p}b^\dag_{p,-}\vert 0\rangle$, where the 
$b^\dag_{p,\sigma}$ are related to $a^\dag_{p,\sigma}$ by the unitary transformation, 
characterised by $\alpha$, that we have already introduced in Eq. (\ref{eq6a}). 
The energy of the state $\vert \alpha \rangle$ can be easily found, and reads
\begin{equation}
\frac{E(\alpha)}{\varepsilon}=-\frac{N}{2}\left(\cos\alpha+\frac{w}{N-1}+\frac{v+w}{2}\sin^2\alpha + 2g\sin\alpha\right)  \ .
\label{eq7}
\end{equation}
By treating $\alpha$ as a variational parameter, one can write 
\begin{equation}
E_{\rm HF} = {\rm min}_\alpha\  E(\alpha).
\end{equation}
This minimum energy, associated with the 
HF ground state, is found (cf. left panel of Fig. \ref{fig2}) for $\alpha = 0$ if 
$v+w<1$ (gray shaded area), and for $\cos\alpha=1/(v+w)$ if $w+v>1$, in the LMG model ($g=0$). 
Instead, for the case of the ELMG model (cf. right panel of Fig. \ref{fig2}), 
the HF ground-state energy
evolves as a function of the parameters in a continuous way.
The symmetry of the curve in the left panel with respect to changing 
the sign of $\alpha$, 
and the existence of a degenerate ground-state for $w+v>1$ in the LMG model, 
can be easily seen from Eq. 
(\ref{eq7}). One could notice that if $\cos \alpha = \phi^2$, the energy 
given by (\ref{eq7}) with $g=0$ assumes the famous ``Mexican hat'' shape, and recognise
the mechanism for spontaneous symmetry breaking. On the other hand, in 
the ELMG model with $g\neq 0$, the symmetry with respect to changing the
sign of $\alpha$ is explicitly broken.
Thus, the ELMG model shows two different types of {\it transitions}: one with $g=0$ (LMG model) between a non-degenerate ground state with $v+w<1$ and a degenerate ground state with $v+w>1$; the other, between a degenerate ground state with $g=0$ and $v+w>1$ to a non-degenerate ground state with $g\ne 0$ and $\forall$ $v$, $w$.

From our discussion, one realizes that the HF approximation for the ground state 
works much better for {\it deformed} states in both models. 
We stress that we obviously use $\alpha$ as a deformation parameter, namely deformed
solution means here $\alpha \ne 0$.
In the lower panels of Fig. \ref{fig1} one sees in fact that, the better the exact 
solution of the LMG model is described in HF, the larger the value of the parameters 
($v$ and $w$) is. This, in turn, corresponds to a larger value of $\alpha$ at 
the minimum of $E(\alpha)$. 
In nuclear physics, there is a widespread tendency among practioners to advocate 
that mean-field methods work better for deformed systems (this may stem, intuitively,
from the idea that a deformed system has already absorbed many correlations 
that would impair the effectiveness of mean-field). 
Although transferring our present conclusions to realistic nuclear 
deformation would require further investigation, 
we still think that our work highlights the appropriateness of mean-field
methods for deformed systems in a sound and pedagogical manner. In particular, 
the comparison between the ELMG Hamiltonian and its LMG limit appears 
to serve this purpose quite well. In this attempt to connect the ELMG model to 
the structure of the nuclear ground-state, we can remind that the matrix elements $-G$ in Eq. (\ref{eq4}) may
be seen as a proxy for the interaction with nucleons outside closed shells that
polarize the system.

In summary, we have presented a new exactly solvable model (ELMG), 
inspired by the LMG model, that has been shown to be useful for applications 
in different fields in physics. The spectrum of applicability of the new model is, 
in principle, broader than that of the LMG model. The new term that has been introduced 
in the ELMG Hamiltonian is responsible for an explicit symmetry breaking leading, 
in general, to a non-degenerate HF ground state, characterised by a variational 
parameter $\alpha$ that is associated with a the description of the system 
in terms of a superposition of bare particles in the upper and lower levels. 
This new term in the Hamiltonian also enables other important differences: 
for instance,  
the coupling between 1$p$-1$h$ and 2$p$-2$h$ is non zero at variance with what happens 
for the LMG in SRPA calculations. We have shown that many-body approximations 
of common use, such as the HF, RPA or SRPA, describe remarkably well the exact 
ELMG solution for the ground state and the first two excited states. 
This shows in a quite transparent manner that deformed systems are amenable 
to a mean-field description. Mean-field methods should be able to provide 
a simple and yet reliable description of complicated real physical systems
when similar conditions are fulfilled. The case of real systems desevres, evidently,
further consideration.

\vspace{1cm}

The authors thank H. Sagawa and K. Hagino for useful discussions. Funding from the European Union's Horizon 2020 research and innovation programme under grant agreement No 654002 is acknowledged.

\section*{Appendix}

We start from the general two-body Hamiltonian of Eq. (\ref{eq1}) and we focus on its interaction term,
\begin{equation}
{\mathcal V} = 
\frac{1}{2}\sum_{\sigma_1,\sigma_2,\sigma_3,\sigma_4}\sum_{p,p^\prime} 
\mathcal{V}_{\sigma_1\sigma_2,\sigma_3\sigma_4} a^\dag_{p\sigma_1}a^\dag_{p^\prime\sigma_2}a_{p^\prime\sigma_4}a_{p\sigma_3}.
\end{equation}
As written in the main text, we write explicitly the various terms corresponding to the two values of $\sigma_i = \pm$ in the sum. 
The results reads 
\begin{eqnarray}
{\cal V} = &\frac{V_{++,--}}{2} \left( J_+^2 + J_-^2 \right) + \frac{V_{+-,-+}}{2} \left(J_+ J_- + J_-J_+ - J_{0-} - J_{0+}\right) 
\nonumber \\
&+ V_{+-,+-}J_{0+} J_{0-}  \nonumber \\
&+ \frac{V_{--,+-}}{2} \left(J_-J_{0-} + J_{0-}J_+\right) + \frac{V_{--,-+}}{2} \left(J_-J_{0-} + J_{0-}J_+\right) 
\nonumber \\
&+ \frac{V_{++,-+}}{2} \left(J_{+}J_{0+} +J_{0+} J_{-} \right) + \frac{V_{++,+-}}{2} \left(J_+J_{0+} + J_{0+}J_-\right) 
\nonumber \\
&+\frac{V_{++,++}}{2} J_{0+}^2 + \frac{V_{--,--}}{2} J_{0-}^2
-\frac{V_{++,++}}{2} J_{0+} - \frac{V_{--,--}}{2} J_{0-}, 
\end{eqnarray}
where we have the introduced two additional operators with respect to those in Eq. (\ref{eq3}), namely: 
\begin{equation}
  J_{0+} = \sum_p a_{p,+}^{\dagger} a_{p,+}, ~~~~J_{0-} = \sum_p a_{p,-}^{\dagger} a_{p,-}.
\end{equation}
These latter operators satsify the commutation relations
\begin{equation}\label{SLMG_newoperators_commutation_rel}
  \left[ J_{0\pm}, J_+ \right] = \pm J_+, ~~~~\left[ J_{0\pm}, J_- \right] = \mp J_-.
\end{equation}
We can write $V_{++,--}$ and $V_{+-,-+}$ as $-V$ and $-W$, following the original LMG model. Since the interaction must be symmetric under 
the exchange of particles, ${\mathcal V}$ takes the following compact form:
\begin{eqnarray}
{\mathcal V} = - &\frac{V}{2} \left( J_+^2 + J_-^2 \right) - \frac{W}{2} \left( \{J_+, J_-\} - J_{0-} - J_{0+})\right)+ \nonumber \\
&+ V_{+-,+-}J_{0+} J_{0-}  \nonumber \\
&+ V_{--,+-} \left( J_-J_{0-} + J_{0-}J_+\right) \nonumber \\
&+ V_{++,-+} \left(J_{+}J_{0+} + J_{0+} J_{-}\right) \nonumber \\
&+\frac{V_{++,++}}{2} J_{0+}^2 + \frac{V_{--,--}}{2} J_{0-}^2
-\frac{V_{++,++}}{2} J_{0+} - \frac{V_{--,--}}{2} J_{0-}. \nonumber \\
\end{eqnarray}
This expression is still difficult to handle because of the presence of operators $J_{0\pm}$. Therefore, we assume the validity of the following approximation
which is also briefly discussed in the main text:
\begin{eqnarray}
V_{+-,+-} &=& V_{++,++} = V_{--,--} = -F \\
V_{--,+-} &=& V_{++,-+} = -G.
\end{eqnarray}
This leads to
\begin{eqnarray}
{\mathcal V} = &-\frac{V}{2} \left( J_+^2 + J_-^2 \right) - \frac{W}{2} \left( \{J_+, J_-\}- \left(J_{0-} + J_{0+}\right)\right) \nonumber \\
&- \frac{F}{2}\left(J_{0+} + J_{0-} \right)^2  
+ \frac{F}{2} \left( J_{0+} + J_{0-} \right) \nonumber \\
&-G \left( J_-J_{0-}+J_{0-}J_+ + J_+J_{0+}+J_{0+}J_- \right).
\end{eqnarray}
The term proportional to $G$ can be rewritten by 
exploiting the commutation relations in Eq. 
(\ref{SLMG_newoperators_commutation_rel}), and one obtains
\begin{eqnarray}
{\mathcal V} & = & -\frac{V}{2} \left( J_+^2 + J_-^2 \right) 
- \frac{W}{2} \left(\{J_+, J_-\} - \left(J_{0-} + J_{0+}\right)\right) 
\nonumber
\\ 
&&- \frac{F}{2}\left(J_{0+} + J_{0-} \right)^2  
+\frac{F}{2} \left( J_{0+} + J_{0-} \right) 
-G ( J_+ + J_-)(J_{0+}+J_{0-}-1). \nonumber \\
\end{eqnarray}
Since the sum of $J_{0+}$ and $J_{0-}$ is the number operator, 
the total Hamiltonian can be written in
the form of Eq.~(\ref{eq2}).

\end{CJK}

\section*{References}
\bibliography{bibliography}

\end{document}